# Reconfigurable metasurface hologram by utilizing addressable dynamic pixels

TIANYOU LI,[1] QUNSHUO WEI,[1] BERNHARD REINEKE,[2] FELICITAS WALTER,[2] YONGTIAN WANG,[1] THOMAS ZENTGRAF,[2] LINGLING HUANG[1*]

[1]*School of Optics and Photonics, Beijing Institute of Technology, Beijing, 100081, China*
[2]*Department of Physics, Paderborn University, Warburger Str. 100, Paderborn, 33098, Germany*
*\*huanglingling@bit.edu.cn*

**Abstract:** The flatness, compactness and high-capacity data storage capability make metasurfaces well-suited for holographic information recording and generation. However, most of the metasurface holograms are static, not allowing a dynamic modification of the phase profile after fabrication. Here, we propose and demonstrate a dynamic metasurface hologram by utilizing hierarchical reaction kinetics of magnesium upon a hydrogenation/dehydrogenation process. The metasurface is composed of composite gold/magnesium V-shaped nanoantennas as building blocks, leading to a reconfigurable phase profile in a hydrogen/oxygen environment. We have developed an iterative hologram algorithm based on the Fidoc method to build up a quantified phase relation, which allows the reconfigurable phase profile to reshape the reconstructed image. Such a strategy introduces actively controllable dynamic pixels through a hydrogen-regulated chemical process, showing unprecedented potentials for optical encryption, information processing and dynamic holographic image alteration.

## 1. Introduction

With tailored artificial subwavelength building blocks, metasurfaces are able to manipulate the local and far-field light, thus enabling a new generation of ultrathin optical elements with a broad range of functionalities [1–8]. By utilizing metasurfaces as information carriers, metasurface holograms are able to store and reconstruct full wave information of target objects. Therefore, they are used increasingly for 3D-displays, data storage, optical metrology and beam shaping applications [9]. However, most of the metasurface holograms are passive devices. That is, their geometric, material and optical properties are fixed once the metasurface is fabricated, which significantly hinders dynamic applications.

Reconfigurable metasurfaces have been proposed to overcome these constraints, owing to their extraordinary ability to dynamically tailor the optical properties, which opens the way to develop dynamic meta-devices. The key for implementing an active metasurface is to explore mechanisms of tuneable hybrid metasurface systems by the incorporation of active materials that change their properties under the proper external stimulus. Ultimately, this will empower dynamic capabilities for polarization control, sensing or imaging [10–12]. Notably, phase change materials and 2D materials, such as germanium-antimony-telluride, graphene, and black phosphorus, are some of the most common ingredients for the realization of such reconfigurable meta-devices [13,14]. By applying a proper optical or thermal excitation, versatile functionalities like switchable dichroism and dynamic plasmonic color displays have been demonstrated [15,16]. In recent years, several attempts have been reported to achieve dynamic metasurface holograms [17,18]. One notable way is to use a mechanically stretchable substrate [17]. With predefined multiplexed holographic images at separate reconstruction planes, the metasurface can dynamically switch the reconstructed images at a specific distance upon stretching. In addition, fully electrically programmable metasurface holograms were realized in the microwave range [18]. However, such functionality cannot easily be achieved at optical wavelength due to the much smaller plasmonic elements and smaller tuning ranges of the optical properties by electric fields.

The recently emerged technology of magnesium-based active plasmonic devices provides a potential solution for reconfigurable optical elements. Magnesium (Mg) exhibits excellent metallic properties at optical frequencies. More importantly, Mg has emerged as an active material once it is hydrogenated towards dielectric magnesium-hydride ($MgH_2$) or dehydrogenated back Mg by loading oxygen, yielding high contrast in its optical properties [19]. Therefore, the hydrogenation/dehydrogenation kinetics of Mg nanoantennas are ideally suited for creating dynamic plasmonic systems [20–22]. Recently, Li et al. demonstrated this principle by building a Mg/Au based holographic metasurfaces. The metasurface consisted of spatially interleaved arrays of Mg and Gold nanoantennas. By hydrogenation/dehydrogenation, the Mg nanoantennas changed their optical properties and, hence, the encoded hologram can be switched [23]. However, this design utilized Mg nanoantennas as independent building blocks and required spatial multiplexing of the holographic information within the metasurface, which unavoidably reduces the available space-bandwidth product for the hologram.

Here, we propose and demonstrate a switchable metasurface phase-only hologram by using the hydrogenation/dehydrogenation process of composite catalytic magnesium-gold-nanoantennas. Our metasurface consists of pure gold V-shaped antennas (PVAs) with fixed scattering phase and composite magnesium-gold V-shaped antennas (CVAs) with switchable scattering phase by the hydrogenation/dehydrogenation. To calculate the phase profile of the holograms, we use an iterative computer-generated hologram (CGH) algorithm. The algorithm generates two target phase profiles with quantified phase relationship for a number of dynamic pixels, which are containing the CVAs. Thus, one holographic image can be reconstructed from the metasurface before the hydrogenation reaction takes place and one after. Through the hydrogenation of the CVAs, the additional holographic information can be read out, which is indiscernible before the hydrogenation or by other visual methods. For switching back to the original holographic information, the CVAs can be dehydrogenated in an oxygen atmosphere, turning the Mg back to its metallic phase.

## 2. Results

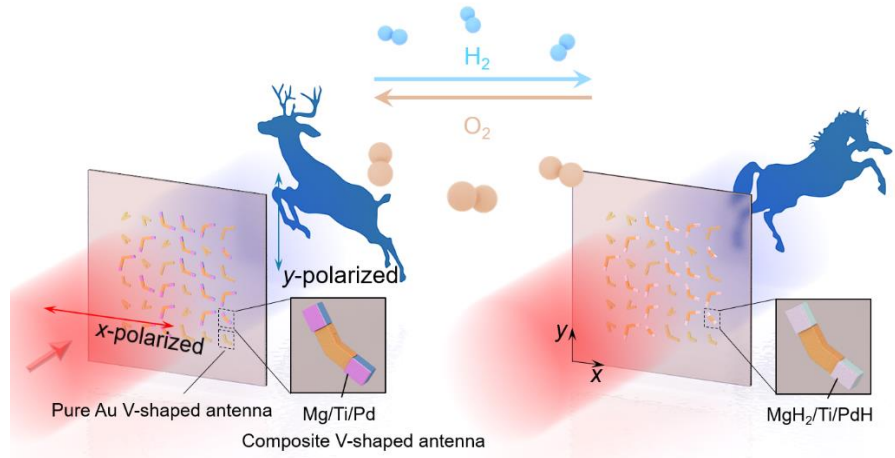

Figure 1. Schematic illustration of the switchable metasurface hologram by utilizing hydrogenation and dehydrogenation processes. The metasurface is composed of a mixture of PVAs and CVAs. After hydrogenation/dehydrogenation, the scattering phase of the CVAs changes, leading to a different reconstructed image.

In the following, we discuss the reconfigurable metasurface approach that is based on composite Au/Mg antennas. Our approach enables dynamical switching between two different holographic images, which is triggered by the surrounding atmosphere of the hologram. Figure 1 schematically illustrates the hybrid design and the working principle. The metasurface can reconstruct two different holograms while residing in air or in hydrogen. In our example, one hologram reconstructs an image of a deer while the other hologram reconstructs an image of a horse. The switching between the two holograms relies on the ability of Mg to form a metallic hydride, which results in a chemical transition of the CVAs' optical properties. Hence, the composite Au/Mg antennas function as dynamic pixels that provide a switchable scattered phase, while the PVAs' scattered phase remains unchanged under different chemical environments. Such a combination of CVAs and PVAs can result in switching the entire phase profile, leading to a dynamic reconstruction of different images. The required phase distribution of such a switchable metasurface hologram is determined by a specially designed holographic algorithm, as described in section 2.3.

## 2.1. Optical properties of composite antennas

Our design is based on a composite V-shaped antenna approach, as shown in Fig. 2(a). The CVAs consist of Au V-shaped antennas (Au arms) together with Mg patches as extensions at both ends. All of the PVAs and CVAs have a 40 nm thickness, and a 10-nm-thick palladium (Pd) catalytic layer is coated on the Mg patch to accelerate the hydrogenation and the dehydrogenation processes [19]. To avoid the alloying of Mg and Pd as well as to release mechanical stress during the hydrogenation process, an additional 5-nm-thick titanium (Ti) layer placed between the Mg and Pd. Note that the PVAs can be considered as CVAs with Mg patch length of zero. The antenna arrays are placed on an indium-tin-oxide (ITO) coated silica glass ($SiO_2$) substrate to ease the two-step electron beam lithography process.

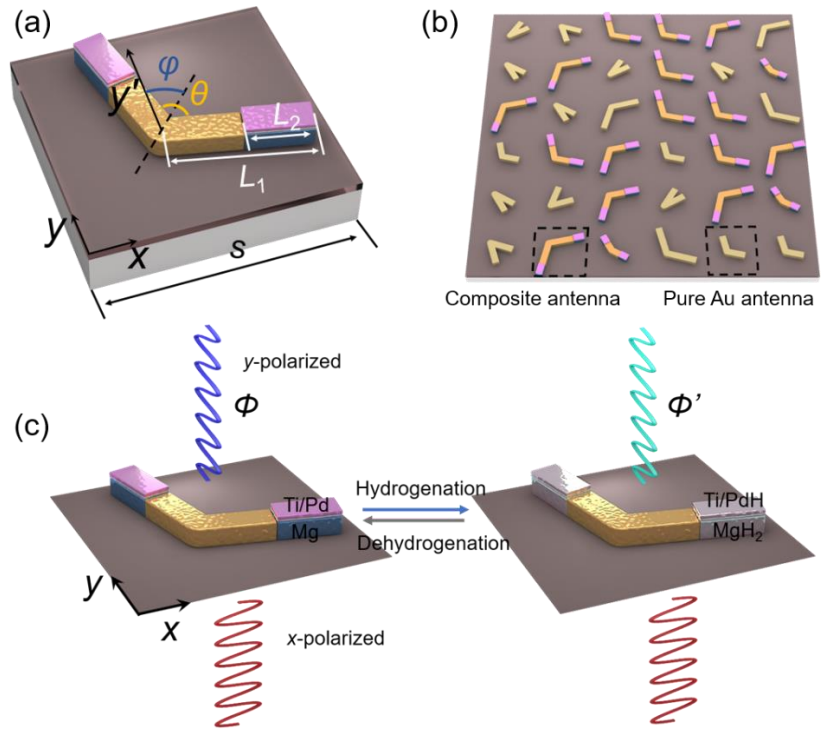

Figure 2. Design principle of the metasurface. (a) Schematic illustration of a single composite V-shaped antenna (CVA) composed of pure Au arms and Mg/Ti/Pd patches. A pure Au antenna

(PVA) can be considered as a composite antenna with $L_2$=0. The dashed line represents the symmetry axis of the V-shaped antenna. $\theta$ is the intersection angle, $\varphi$ is the azimuthal angle with respect to the symmetry axis of the V-shaped antenna, and y' is the local coordinate, which is parallel to the global y-axis. (b) Illustration of the metasurface configuration: half of the pixels are made of CVAs that function as dynamic pixels with switchable scattering properties, while the other half is made of PVAs that function as static pixels. (c) Hydrogenation/dehydrogenation of the metasurface results in a modified phase of the scattered field. After hydrogenation, the scattered field of the composite antenna changes from $\Phi$ to $\Phi'$, and vice versa by utilizing $O_2$ for the dehydrogenation process.

Figure 2(b) illustrates the arrangement of the V-shaped antennas to form a switchable phase-only metasurface hologram. The metasurface is composed of the same number of randomly distributed PVAs and CVAs. The geometric sizes of the CVAs and PVAs are designed to cover the entire 0-2π scattering phase range (for details see Sec. 2.2). Since the PVAs remain their optical properties during the hydrogenation process, we further investigate the phase modulation properties of the CVAs, as shown in Fig. 2(c). In ambient air, Mg shows a plasmonic behavior, which means that the Au nanoantennas with the Mg patches (Au/Mg) act as single composite structures with plasmonic resonances that generate a scattering field $\Phi$. In contrast, in a hydrogen environment, the Mg changes to $MgH_2$ and loses its plasmonic properties. Hence, the effective lengths of the Au/$MgH_2$-composite antennas are reduced and the resonance frequencies are shifted, resulting in a changed scattering field $\Phi'$. By designing a set of V-shaped Au/Mg-composite antennas with a constant phase shift when switching to Au/$MgH_2$ while keeping a uniform scattering amplitude results in a different phase pattern for the hologram. Note that through a dehydrogenation process utilizing oxygen, the $MgH_2$ can be reversed back to Mg with its metallic properties.

## 2.2 Nanoantenna design

V-shaped plasmonic nanoantennas are suitable building blocks for holographic metasurfaces. Their optical properties can be tuned by their geometric size and shape to obtain a desired scattering of light at the working wavelength of 1200nm. In the case of phase-only holograms, the scattering amplitude should be constant for all nanoantennas and the scattering phase should cover a full 0-2π range. The V-shape has the advantage that both symmetric and anti-symmetric localized surface plasmon polariton modes can be excited by far-field radiation. If the incident light is polarized along or orthogonal to the symmetry axis of the V-shaped antenna, the polarization of the scattered light is the same as of the incident light. This property is related to the excitation of only one plasmon mode, either the symmetric or the anti-symmetric mode. Any other polarization direction for the incidence light excites both plasmon modes simultaneously but with different phases and amplitudes. Hence, the scattered light can have a different polarization, phase and amplitude depending on the incident polarization. In our design, we set the symmetry axis of V-antennas under an azimuthal angle $\theta = 45°$ with respect to the y-axis (Fig. 2(a)). With all degrees of freedom of the antenna geometry (arm length $L_1$, patch length $L_2$, and intersection angle $\theta$), the phase and amplitude of the cross-polarized scattered light can be designed within a large parameter space for a given excitation wavelength $\lambda_0$. To cover a phase range of 0-2π, the antenna design can be mirrored for an additional phase shift of π [24].

We used rigorous coupled wave analysis (RCWA) to analyze the scattering field of the PVAs and CVAs over a wide parameter range to identify suitable antenna geometries that can satisfy the required phase relation. In the simulations, we used periodic boundary conditions with a lattice constant of $d = 0.63$ μm. The refractive indices of Au, Mg, $MgH_2$ are taken from Johnson and Christy [26], Palik [27], and Griessen [28] respectively. The permittivities of Pd and PdH are taken from Rottkay [29]. We note that after the hydrogenation process the additional Pd and Ti layers on top of the extended patches still have metallic properties. Moreover, the metallic properties of the Mg/Ti/Pd composite patches differ from the optical properties of Au and cannot be modeled by the same material parameters. Thus, CVAs, both before and after hydrogenation, show different scattering properties from the PVAs with corresponding arm length. Hence, the constitute Au antenna part of the CVAs needs to have a different geometry compared to the PVAs to gain the same scattering phase and amplitude. Therefore, we optimize two different sets of antenna arrays, one set with PVAs and the other set with CVAs. For the PVA array, the set of geometries with approximately the same scattering amplitude and a constant phase difference of $\pi/4$ (8 phase-orders) are chosen to obtain entire 0-$2\pi$ coverage (Fig. 3). While for the CVA array, the suitable geometries are selected with more restrictions, since the phase difference for each antenna before and after the hydrogenation process needs to be constant. The selected CVAs both before (upper side) and after (bottom side) hydrogenation are shown in Fig. 3(d). The corresponding scattering amplitudes and phases for these antennas are shown in Fig. 3(e) and Fig. 3(f), respectively. From Fig. 3(f), it is noteworthy that before hydrogenation, when the Mg patches have metallic properties, the scattering phases of the CVAs (blue dotted line) are consistent with that of the PVAs (as shown in Fig. 3(c)). Thus, the condition of a phase-only hologram with uniform amplitude can be satisfied. After the hydrogenation, the scattering phase of all CVAs is uniformly shifted by $\pi/3$. Note that there are slight amplitude differences before and after the hydrogenation process, which need to be carefully considered in the holographic design algorithm. The detailed parameters of the PVAs and the CVAs are listed in Table 1 and Table 2, respectively.

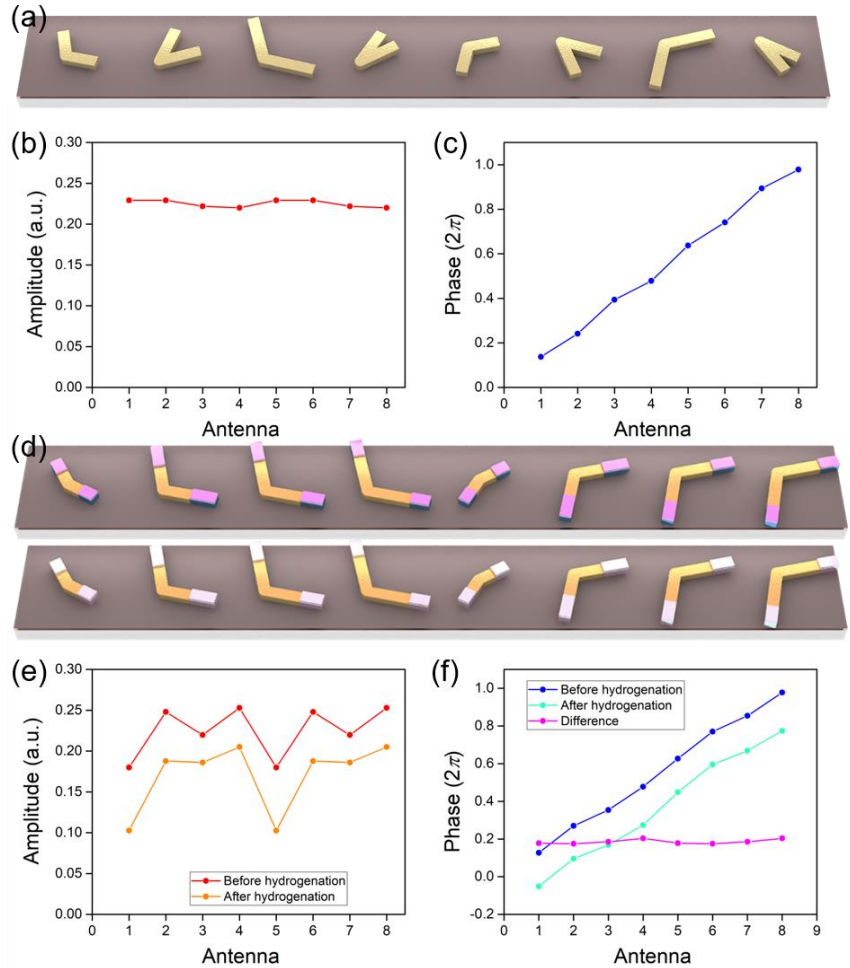

Figure 3. The scattering amplitude and phase of eight selected PVAs and CVAs. (a) PVA array with a phase difference of π/4 corresponding to the antenna 1 to 8 (left to right). (b) and (c) Scattering amplitude and phase of each PVA. (d) CVA array before (upper panel) and after hydrogenation (bottom panel). (e) and (f) Scattering amplitude and phase of each CVA before and after hydrogenation. The pink dotted line indicates the phase difference between the two curves.

Table 1. The selected geometry parameters of the PVAs.

| Antenna | 1 | 2 | 3 | 4 |
|---|---|---|---|---|
| $\theta$ [°] | 115 | 55 | 120 | 30 |
| $L_1$ [µm] | 0.2 | 0.26 | 0.33 | 0.26 |

Table 2. The selected geometry parameters of the CVAs.

| Antenna | 1 | 2 | 3 | 4 |
|---|---|---|---|---|
| $\theta$ [°] | 150 | 115 | 115 | 100 |
| $L_1$ [µm] | 0.2 | 0.34 | 0.37 | 0.39 |
| $L_2$ [µm] | 0.1 | 0.16 | 0.14 | 0.12 |

## 2.3. Design of the holograms

The key to realizing a reconfigurable metasurface is to set quantified phase relations between two holograms. Figure 4 shows a flow chart of the holographic design algorithm that is characterized by an iterative loop between two hologram planes and two reconstruction planes. The algorithm is based on the Fidoc method. Such an algorithm can be used to distribute the encoding noise to an area of less importance ("don't care area") and therefore improves the signal-to-noise ratio [25].

We chose the hologram to be reconstructed in the Fresnel diffraction range at the distance of 20 mm. Therefore, the light propagation during the iteration is simulated by:

$$U_d(x,y) = \frac{\exp(jkd)}{jkd} \exp\left[\frac{jk}{2d}(x^2+y^2)\right] \times \iint \left\{ U_0(x_0,y_0) \exp\left[\frac{jk}{2d}(x_0^2+y_0^2)\right] \right\} \times \exp\left[-j2\pi\left(x_0\frac{x}{\lambda d} + y_0\frac{y}{\lambda d}\right)\right] dx_0 y_0 \quad (1)$$

whereas $U_o$ and $U_d$ are the complex amplitudes in the hologram and the object plane, respectively, k is the wave vector, $(x_0, y_0)$ are the coordinates of the metasurface hologram, and $(x, y)$ are the coordinates of the reconstruction plane. Note that, half of all pixels in the hologram are selected randomly and function as dynamic pixels, which means that a phase shift of $\alpha$ (π/3 in our case by considering all the restrictions as stated in the Supplementary Material) is added or subtracted to the other phase profile. This ensures that the optimized phase profiles of the two holograms have a phase difference of $\alpha$ for the dynamic pixels while the phase of the remaining pixels stays the same. Meanwhile, all scattering amplitudes are optimized to be uniform for such a phase-only hologram. Therefore, by combining the 'dynamic' pixels and 'static' pixels together, we can achieve phase conversion between the two hologram planes which will result in the reconstruction of two totally different holographic images.

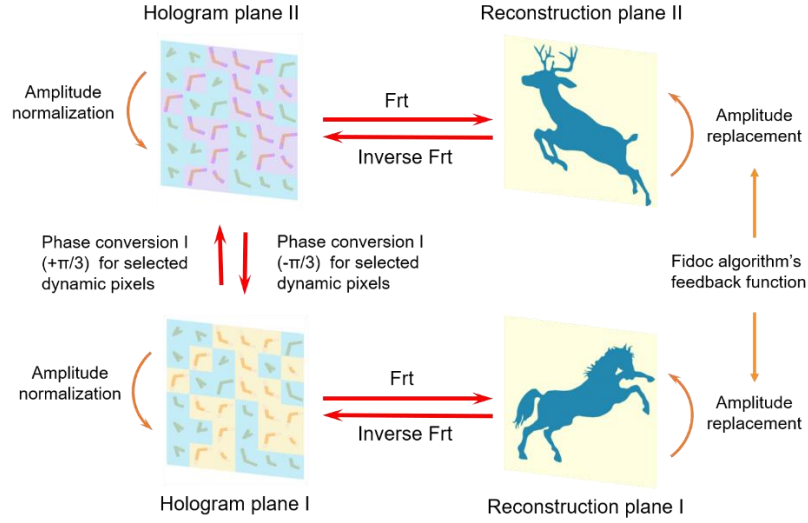

Figure 4. Flow chart of the holographic design algorithm which illustrates the iterative loop between the two hologram and reconstruction planes to form a quantified phase relation. The simulated light propagation between the hologram and the reconstruction plane is performed by a Fresnel transformation (FrT).

## 3. Results and discussion

To demonstrate our dynamic hologram scheme, we select two images (deer and horse) to obtain the encoded phase information. Randomly half of the pixels are designed as dynamic pixels in the algorithm (total 317×317 pixel numbers). The numerical Fresnel reconstruction images of the ideal phase profiles with uniform amplitude distribution are shown in Fig. 5(a) and 5(b). With those dynamic pixels undergoing the $\pi/3$ phase shift by mimicking the hydrogenation process, the reconstruction image should switch from the deer (Fig. 5(a)) to the image of the horse (Fig. 5(b)). Thus we successfully confirm that our hologram algorithm works.

To determine the holographic images from the reconfigurable metasurface, we adopt a method that is combining full-wave simulations and Fresnel diffraction reconstruction. First, we choose the set of PVA and CVA arrays that function as the static and dynamic pixels, respectively, to achieve an 8-step phase encoding of the whole hologram profile. Due to the deviations of the amplitude and phase of the selected V-shape antennas (realistic field profile) to the phase-only profile obtained by the designed algorithm, we then calculated the numerical Fresnel reconstruction of such realistic complex amplitude profile. That is, we use the simulated complex amplitude of the CVAs together with that of the PVAs (as indicated in Fig. 3(e)-3(f)) to calculate the reconstruction image before hydrogenation (Fig. 5(c)). For obtaining the second image, we use the complexed amplitude data after hydrogenation where the Mg patches changed their properties (Fig. 5(d)). The reconstruction images calculated for the metasurface with the V-shaped antennas are consistent with the ideal phase-only reconstruction images, which demonstrates the effectiveness of such a metasurface scheme.

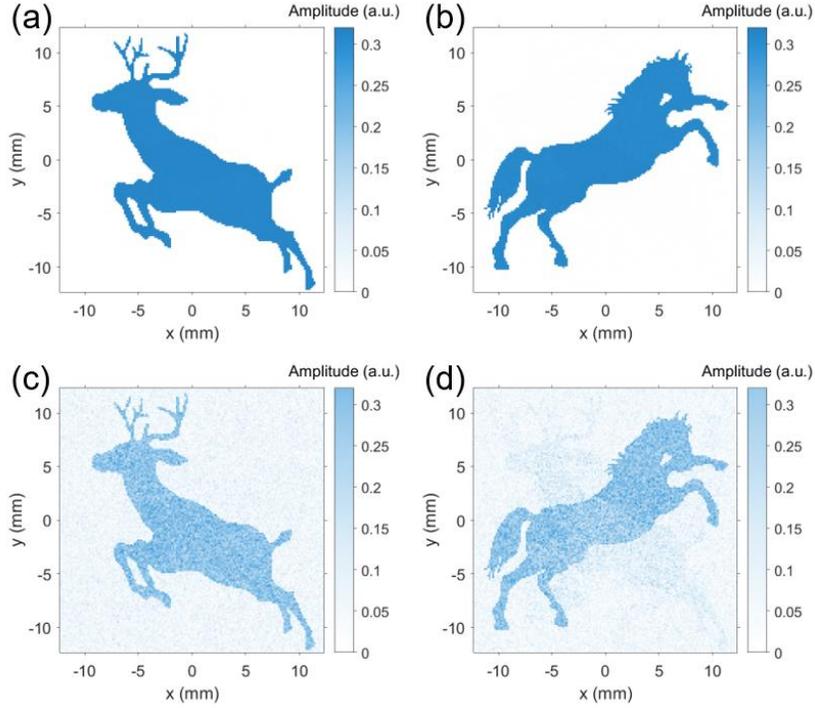

Figure 5. Reconstruction images of the ideal phase distribution and the metasurface hologram. (a) and (b) are the reconstruction images of a calculated ideal phase-only hologram of the deer and the horse, respectively. (c) and (d) are the reconstruction images of the complex amplitude profile of the realistic PVA and CVA arrays, calculated by the Fresnel diffraction algorithm before and after the hydrogenation.

Note that although all the scattering phase of the selected antennas are equally distributed, the scattering amplitudes of the CVAs exhibit considerable fluctuation (see Fig. 3(e)). That is, such reconfigurable metasurface hologram cannot fully satisfy the uniform amplitude condition, which is required for the phase-only holographic scheme. Furthermore, the scattering amplitude of the CVAs drops slightly after hydrogenation, which causes an amplitude contrast between the dynamic (CVAs) and static pixels (PVAs). The amplitude fluctuation results in the appearance of noise in the reconstructed image, which decreases the image quality and intensity, as shown in Fig.5(d). Such an effect is unavoidable since the effective arm lengths of the CVAs are considerably shortened after hydrogenation. However, higher image quality can be obtained by further improving our holographic algorithm considering the realistic complex amplitude (amplitude + phase) modulation in the design rather than the phase-only concept.

The diffraction efficiencies of the two holograms which can reconstruct a deer and horse are 3.7% and 3.2%, respectively. The diffraction efficiency here is defined as the product of the metasurface's transmission efficiency and the efficiency of the holographic reconstructed image. Note that the transmission efficiency of the metasurface can be obtained by weighted transmission energies of the chosen nanoantennas. In addition, to quantitatively estimate the quality of reconstructed images and improve the holographic algorithm, peak signal-to-noise ratios (PSNR) and the correlation coefficients are analyzed as criteria. The PSNRs of the reconstruction images of the complex amplitude profile of the realistic PVA and CVA arrays, calculated by the Fresnel diffraction algorithm before and after the hydrogenation, are 57.5872 and 57.3292, for the deer and horse, respectively. And the correlation coefficients between reconstructed images and their corresponding target objects are 0.9226 and 0.8928, for deer and horse, respectively. These values mean that the qualities of reconstructed images are satisfactory, and the crosstalk between the two images are small.

## 4. Conclusion

In summary, we proposed and theoretically demonstrated a concept to realize an addressable dynamic metasurface hologram by utilizing hierarchical reaction kinetics of magnesium upon hydrogenation/dehydrogenation. A switchable metasurface hologram is obtained by utilizing composite Au/Mg and pure Au V-shaped antennas. We realized a smart holographic algorithm to build quantified phase relations that allows the switchable scattering phase for selected dynamic pixels to reshape the reconstruction image. The metasurface hologram can be used in a hydrogen or oxygen environment, which enables a promising platform to achieve dynamic functionalities. Our proposed method provides a new platform for active metasurface phase profiles, implying potential applications in optical encryption, information processing, smart sensors, and dynamic meta-device systems.

**Funding**

The authors acknowledge the funding provided by the National Key R&D Program of China (No. 2017YFB1002900) and the European Research Council (ERC) under the European Union's Horizon 2020 research and innovation program (grant agreement No. 724306). We also acknowledge the NSFC-DFG joint program (NSFC No. 61861136010, DFG No. ZE953/11-1). L.H. acknowledge the support from National Natural Science Foundation of China (No. 61775019) program, Beijing Municipal Natural Science Foundation (No. 4172057) and Beijing Nova Program (No. Z171100001117047).